\documentclass{amsart}

\usepackage{amsmath, color, dsfont}
\usepackage[utf8]{inputenc}
\usepackage{geometry}
\usepackage{paralist} 
\usepackage{subfig} 
\usepackage[usenames,dvipsnames]{xcolor}
\usepackage{graphicx}
\usepackage{feynmf}
\usepackage[all]{xy}

\usepackage{amsbsy,amssymb,amscd,amsfonts,latexsym,amstext,delarray,
amsmath,graphicx} 
\input xypic
\usepackage{color}

\newtheorem{thm}{Theorem}[section]
\newtheorem{prop}[thm]{Proposition}
\newtheorem{cor}[thm]{Corollary}
\newtheorem{lem}[thm]{Lemma}

\newtheorem{defn}[thm]{Definition}
\newtheorem{rem}[thm]{Remark}

\numberwithin{equation}{section}

\def\bS{{\mathbb S}}

\def\C{{\mathbb C}}

\renewcommand{\H}{{\mathbb H}}
\def\N{{\mathbb N}}
\renewcommand{\P}{{\mathbb P}}
\def\Q{{\mathbb Q}}
\def\Z{{\mathbb Z}}
\def\R{{\mathbb R}}

\def\cA{{\mathcal A}}
\def\cB{{\mathcal B}}
\def\cC{{\mathcal C}}

\def\cH{{\mathcal H}}

\def\cK{{\mathcal K}}

\def\cP{{\mathcal P}}

\def\cS{{\mathcal S}}

\def\Tr{{\rm Tr}}

\def\Aut{{\rm Aut}}
\def\Sp{{\rm Spec}}

\def\cancel#1#2{\ooalign{$\hfil#1\mkern1mu/\hfil$\crcr$#1#2$}}
\def\Dirac{\mathpalette\cancel D}
\def\dirac{\mathpalette\cancel \partial}
\def\cutint{{\int \!\!\!\!\!\! -}}

\title[Spectral triples and QSM]{Type III $\sigma$-spectral triples and 
quantum statistical mechanical systems}
\author{Mark Greenfield, Matilde Marcolli, Kevin Teh}
\address{Mathematics Department, Caltech, 1200 E.~California Blvd. Pasadena, CA 91125, USA}
\email{mgreenfi@caltech.edu}
\email{matilde@caltech.edu} 
\email{kkwteh@gmail.com}
\date{}

\begin{document}
\maketitle

\begin{abstract}
Spectral triples and quantum statistical mechanical systems are two
important constructions in noncommutative geometry. In particular, both lead to
interesting reconstruction theorems for a broad range of geometric objects,
including number fields, spin manifolds, graphs. There are similarities between
the two structures, and we show that the notion of type III $\sigma$-spectral triple,
introduced recently by Connes and Moscovici, provides a natural bridge between
them. We investigate explicit examples, related to the Bost--Connes
quantum statistical mechanical system and to Riemann surfaces
and graphs.
\end{abstract}

\section{Introduction}

\subsection{Reconstruction theorems in Noncommutative Geometry}

The prototype model of a {\em reconstruction theorem} in Noncommutative
Geometry is also the motivating origin of the very notion of noncommutative spaces,
namely the Gelfand--Naimark theorem, which shows that it is possible to reconstruct
(up to homeomorphism) a compact Hausdorff topological space $X$ from the commutative
$C^*$-algebra of continuous functions $C(X)$ (up to $C^*$-isomorphism). This reconstruction
theorem, which shows that considering compact Hausdorff topological space is equivalent
to considering commutative $C^*$-algebras motivated the notion of a noncommutative
topological space as the datum of a noncommutative $C^*$-algebra, \cite{Co94}.

More recently, reconstruction theorems have been proved for the analogs, in Noncommutative
Geometry, of compact spin Riemannian manifolds, namely {\em spectral triples}, a notion 
we will discuss more in detail in the rest of this paper. The prototype for this kind of reconstruction
theorem is the reconstruction of compact spin Riemannian manifolds from the data of a commutative
spectral triple satisfying a set of axioms involving appropriate notions of orientability, Poincar\'e
duality and smoothness, \cite{Co08}. For some variants and generalizations see \cite{LRV}, 
\cite{Zhang}. Similar results have been proved for other classes of 
noncommutative spaces, such as almost-commutative geometries, \cite{Cac}. With a different
approach, a reconstruction theorem for Riemannian manifolds based on zeta functions of
spectral triples was proved in \cite{CordJ}, and a complete invariant in Riemannian geometry 
based on noncommutative methods was obtained in \cite{CoCKM}. A reconstruction
theorem for Riemann surfaces up to (anti)conformal equivalence, based on the zeta functions
of a spectral triple, was obtained in \cite{CorMa1}, with an analogous result for graphs in 
\cite{dJ}. 

At the same time, {\em quantum statistical mechanical systems}, another important 
construction widely used in Noncommutative Geometry, have also been used to obtain
reconstruction theorems based on noncommutative methods, most notably the reconstruction
result for number fields obtained in \cite{CorMa2}, and a similar result for graphs in \cite{CorMa3}.

Our purpose in this paper is to present a comparative analysis of these two constructions
in Noncommutative Geometry: {\em spectral triples} and {\em quantum statistical mechanical 
systems}. We show that there are deep conceptual similarities between the two notions,
but also important differences. We investigate to what extent one can transform one type of
structure into the other and extract from both of them similar type of information on the
underlying geometric objects and the related classification problems.

\medskip

The paper is organized as follows. In Sections \ref{S3sec} and \ref{QSMsec} we
review well known material about spectral triples and quantum statistical mechanical
systems, respectively, trying to outline the differences and the apparent similarities
between the two structures. Then in Section \ref{BCsec} we focus on the example
of the Bost--Connes quantum statistical mechanical system \cite{BC} and on a 
related structure based on the Riemann gas with supersymmetry considered in
\cite{Jul}, \cite{Spec1}, \cite{Spec2}. We show that the data of the Bost--Connes
system determine a type III $\sigma$-spectral triple, where the sign $F$ of the
Dirac operator and the twisting automorphism $\sigma$ are determined by the
Liouville function. A similar construction works for the Riemann gas with supersymmetry,
using the M\"obius function instead of the Liouville function. We discuss some properties
of eta functions and KMS states.
In Section \label{Lambdasec} we consider instead the case of limit sets of Schottky
groups and of graphs, where one can use a very similar construction covering both
cases, as in \cite{CorMa1}, \cite{CorMa3}, \cite{dJ}. We show that the 
spectral triples considered in the reconstruction results of \cite{CorMa1} and \cite{dJ}
for Riemann surfaces and graphs have some undesirable features, such as the
fact of not admitting an interesting sign and compatible even structure. We propose
then a construction of a type III $\sigma$-spectral triple for these cases, analogous
to the construction described for the Bost--Connes case, that starts from the
quantum statistical mechanical systems of \cite{CorMa3}, \cite{Lott}. 
 
\section{Spectral triples} \label{S3sec}

This section gives a review of well known notions about spectral triples
in Noncommutative Geometry. For more details, we refer the reader to
\cite{Co94}, \cite{CoS3}, \cite{CoMa-garden}, \cite{CoMo}.

The notion of a {\em spectral triple} \cite{CoS3} provides an abstract
algebraic version of the main property satisfied by a compact Riemannian
spin manifold. The main purpose of this formulation is to encode the
data of the manifold and its spin and metric structures in terms that avoid
the use of {\em local charts}, which does not have a direct analog in the
noncommutative setting, and instead focuses on the use of operator
relations in a Hilbert space setting. More precisely, we recall the following
well known definition.

\begin{defn}\label{S3def}
A spectral triple  $({\mathcal A},{\mathcal H},D)$
consist of the following data:
\begin{itemize}
\item an involutive algebra $\cA$;
\item a representation $\pi:\cA \to \cB(\cH)$ in the algebra $\cB(\cH)$ of bounded
linear operators on a separable Hilbert space $\cH$;
\item a self adjoint operator $D=D^*$ acting on $\cH$ with a dense domain 
${\rm Dom}(D)\subset \cH$;
\item the condition that $D$ has compact resolvent, namely that $(1+D^2)^{-1/2}$ 
belongs to the algebra $\cK \subset \cB(\cH)$ of compact operators;
\item the condition that all commutators $[\pi(a),D]$ of elements in the algebra, $a\in \cA$, 
with the Dirac operator $D$ are bounded operators, $[\pi(a),D]\in \cB(\cH)$.
\end{itemize}
The spectral triples is said to be {\em even} if there is a $\Z/2$- grading 
$\gamma$ on $\cH$ with the property that 
$$ [\gamma,\pi(a)]=0, \ \forall a\in \cA,  \ \ \ \text{ and } \ \ \   D\gamma =-\gamma D. $$
\end{defn} 

\subsection{Spectral triples and Spin geometry}

One can see that the structure described in Definition \ref{S3def} reflects the
main properties of a compact smooth Riemannian spin manifold $M$, by considering
the case where the involutive algebra $\cA=\cC^\infty(M)$ is the algebra of smooth
complex valued functions on $M$ with the involution given by complex conjugation
and where the Hilbert space is $\cH=L^2(M,\bS)$, the space of square-integrable
sections of the spinor bundle $\bS$ of $M$ (with respect to a choice of spin structure),
on which elements of $\cC^\infty(M)$ act as multiplication operators.
One can take the operator $D=\dirac_M$ to be the Dirac operator acting
on sections of the spinor bundle. One can see easily that these data satisfy all the
conditions of Definition \ref{S3def}. For example, the commutator of the Dirac
operator with a smooth function is given by $[\dirac_M,f ]=c(df)$, the differential $df$ acting
on spinors by Clifford multiplication. The data $(C^\infty(M),L^2(M,\bS),\dirac_M)$
are usually referred to as the canonical spectral triple of a Riemannian spin
manifold. It is well known that, for example, the Riemannian metric on $M$ is
encoded in the data of the spectral triple and can be recovered using the
formula for the distance, \cite{Co-hyp},
$$ d(x,y)= \sup \{ |f(x) - f(y)| \,\,: \,\, f\in C^\infty(M), \,\, \| [\dirac_M,f] \| \leq 1 \}. $$

\subsection{Spectral metric geometry and the Dirac sign}

A Dirac operator $D$ as in Definition \ref{S3def} has a polar decomposition
$D =|D|\, F$, into a positive part and a sign operator, $F^2 =1$.
In the manifold case, the sign $F$ is non-trivial, and in fact it carries important
geometric information, since it represents the fundamental class of $M$ in $K$-homology.
(For a detailed treatment  of these topics see for instance \cite{HR}.) However, in
Noncommutative Geometry, one sometimes considers also a wider class of
spectral triples where the sign $F$ may be trivial. The triviality of the sign causes
problems when one wishes to use the spectral triple setting to apply index theory
techniques, but such spectral triples are still useful, for instance because they
can still define {\em quantum metric spaces} in the sense of \cite{Rieff}, 
and also because they can carry interesting families of zeta functions.

\subsection{Summability, dimensions, and zeta functions}

Zeta functions for spectral triples are defined under a condition of {\em finite
summability}, which we recall in the following.

\begin{defn}\label{S3finsum}
A spectral triple $({\mathcal A},{\mathcal H},D)$ is finitely summable if 
for $\beta >>0$ sufficiently large the operator $|D|^{-\beta}$ is of trace class,
\begin{equation}\label{finsumD}
 \Tr (|D|^{-\beta}) <\infty. 
\end{equation}
The smallest $\beta_c\geq 0$ such that \eqref{finsumD} holds for all $\beta>\beta_c$
is called the {\em metric dimension} of the spectral triple. 
A spectral triple $({\mathcal A},{\mathcal H},D)$ is $\theta$-summable
if for all $t>0$ the heat kernel operator $\exp(-t D^2)$ is of trace class, 
\begin{equation}\label{thetasumD}
 \Tr (e^{-t D^2}) <\infty. 
\end{equation}
\end{defn}

In the case of a finitely summable spectral triple, one can define a family of
zeta functions associated to the data $({\mathcal A},{\mathcal H},D)$ of the
spectral triple and defined as
\begin{equation}\label{zetaaD}
\zeta_a(s) = \Tr(\pi(a) \, |D|^{-s}), 
\end{equation}
where $s$ is a complex variable, and $a\in \cA$, acting on $\cH$ through
the representation $\pi$. One also considers similar zeta functions of the form
$\Tr(\pi(b)|D|^{-s})$, where $b$ is an element of the algebra $\cB$ generated by $\delta^k(\pi(a))$
and $\delta^k([D,\pi(a)])$, for $a\in \cA$ and $\delta(-)=[|D|,-]$.

The zeta functions obtained in this way provide a refined notion of dimension
for noncommutative spaces, given by the {\em dimension spectrum}.

\begin{defn}\label{dimSp}
The dimension spectrum of a finitely summable spectral triple $({\mathcal A},{\mathcal H},D)$
is the subset $\Sigma = {\rm DimSp}({\mathcal A},{\mathcal H},D) \subset \C$ given by the set 
of poles of the zeta functions $\zeta_b(s)$, for $b\in \cB$.
\end{defn}

Since points of the dimension spectrum correspond to poles of the zeta functions of
the spectral triple, one can consider the associated residues,
\begin{equation}\label{ResZetas}
{\rm Res}_{s=\beta \in \Sigma} \,\, \zeta_a(s) =: \cutint a\, D^{-\beta},
\end{equation}
where the meaning of the notation on the right-hand-side of \eqref{ResZetas} is that
one interprets these resides as integration in dimension $\alpha \in \Sigma$, where
$\Sigma$ is the dimension spectrum.

\subsection{Zeta functions and eta functions} 

As we have seen above, for a finitely summable spectral triple,
the zeta function of the Dirac operator $D$ is defined as $\zeta_D(s)=\Tr(|D|^{-s})$.
In Riemannian geometry, another very useful series associated to a self-adjoint
operator of Dirac type is the {\em eta function} $\eta_D(s)$ that measures spectral 
asymmetry and that gives rise to the {\em eta invariant}, as the special value $\eta_D(0)$.
The eta function of a Dirac operator with trivial kernel is defined as
\begin{equation}\label{etaD}
\eta_D(s) = \Tr(F \, |D|^{-s}) = \sum_{\lambda \in \Sp(D)} {\rm sign}(\lambda)\, |\lambda|^{-s}.
\end{equation}

Similarly to what is usually done with zeta functions of spectral triples, one can
consider a family of eta functions of the form
\begin{equation}\label{etaD}
\eta_{a,D}(s) = \Tr(F \,a\, |D|^{-s}), \ \ \text{ for $a$ in $\cA^\infty$}.
\end{equation}

\subsection{Eta functions on manifolds}

In the case of the eta function of a classical self-adjoint elliptic first order differential operator
$\Dirac$ on a manifold of dimension $n$, the eta function can be expressed as, \cite{Gil},
$$ \eta_{\Dirac}(s) = \sum_{\lambda\in {\rm Spec}(\Dirac)\smallsetminus \{ 0 \}} \lambda
\, |\lambda|^{-s-1} = 
\frac{1}{\Gamma(\frac{s+1}{2})} \int_0^\infty t^{\frac{s+1}{2}-1} (\sum \lambda e^{-t \lambda^2}) \, dt $$
$$ = \frac{1}{\Gamma(\frac{s+1}{2})} \int_0^\infty t^{\frac{s+1}{2}-1} \Tr(\Dirac\, e^{-t \Dirac^2})\, dt. $$
In turn, the trace $\Tr(\Dirac\, e^{-t \Dirac^2})$ has an asymptotic expansion from the heat kernel
asymptotics, 
\begin{equation}\label{etaheat}
 \Tr(\Dirac\, e^{-t \Dirac^2}) \sim \sum_{k=0}^N  c_k(\Dirac) \, t^{\frac{k-n-1}{2}}, 
\end{equation} 
which in turn gives
\begin{equation}\label{etapoles}
 \eta_{\Dirac}(s) \sim \frac{1}{\Gamma(\frac{s+1}{2})} \sum_{k=0}^N \frac{2 c_k(\Dirac)}{s+k-n}, 
\end{equation} 
so that one finds poles at $s=n-k$ with residues
$$ {\rm Res}_{s=n-k} \eta_{\Dirac}(s) = \frac{2 c_k(\Dirac)}{\Gamma(\frac{s+1}{2})}. $$

In the general setting of noncommutative geometry, one does not necessarily expect to
always  have a similar a similar form of the heat kernel asymptotics. However, useful 
forms of heat kernel asymptotics exist and are related to asymptotics for the spectral
action functional, \cite{CC1}.

\subsection{The spectral action}

It is customary to consider an {\em action functional} for finitely summable spectral
triples, given by the {\em spectral action}, introduced in \cite{CC1}. The functional
is defined as
\begin{equation}\label{SpAct}
 \Tr (f(D/\Lambda)), 
\end{equation}
where $D$ is the Dirac operator of the spectral triple, 
$\Lambda$ is a mass scale that makes $D/\Lambda$ dimensionless, and 
$f >0$ is an even smooth function, which is usually taken to be a smooth
approximation of a cutoff function. When written in this form, one considers the
argument of this functional to be the Dirac operator $D$, so that the space of {\em fields}
of the corresponding field theory would be the space of all possible Dirac operators $D$
compatible with fixed data $(\cA,\cH)$, so that $(\cA,\cH,D)$ is a spectral triple, possibly
with additional constraints. A variant of \eqref{SpAct} consists of allowing twists of the
Dirac operator by ``gauge potentials". This means replacing \eqref{SpAct} with the
similar expression 
\begin{equation}\label{SpAct2}
 \Tr (f(D_A/\Lambda)), 
\end{equation}
where $D_A = D+A$, where the gauge potential $A$ is a finite sum 
$A=\sum_i \pi(a_i)\, [D,\pi(b_i)]$ with $a_i, b_i \in \cA$.

It was shown in \cite{CC1} that  there is an
asymptotic expansion for $\Lambda\to \infty$ of the spectral action
functional of the form
\begin{equation}\label{SpActAsympt}
 \Tr (f(D/\Lambda))\sim\,\sum_{k \in \Sigma^+}\,f_{k}\,\Lambda
^{k}\,{\int\!\!\!\!\!\!-}\,|D|^{-k} +\,f(0)\,\zeta_{D}(0)+\,o(1),
\end{equation}
where $f_{k}=\,\int_{0}^{\infty}f(v)\,v^{k-1}\,dv$ are the momenta of $f$ and
where the summation runs over the positive part of the dimension spectrum 
$\Sigma^+\subset \Sigma$. In the case of manifolds and of almost commutative
geometries, the residues \eqref{ResZetas} that appear in this asymptotic
expansion have a description in terms of local expressions in the curvature
tensors of the Riemannian metric and in the curvatures of the gauge potentials,
that recover some of the classical action functionals of physics, including the
Einstein--Hilbert action of General Relativity, the Yang--Mills action functionals
of gauge theories, and for a suitable choice of the almost commutative geometry
the bosonic part of the Lagrangian of the Standard Model of elementary particle
physics, see \cite{CCM} and Chapter 1 of \cite{CoMa-book}.

\subsection{Spectral action, summability, and spectral asymmetry}

One usually considers the spectral action in the finitely summable case,
and especially for the almost commutative geometries used in models of
gravity coupled to matter. However, one can consider the spectral action
functional in more general cases.

For example, in the case of a $\theta$-summable $D$, one can consider
the test function $f(t)=\exp(-t^2)$ and the spectral action
$$ \Tr(f(D/\Lambda))= \Tr(e^{-\frac{1}{\Lambda^2} D^2}) < \infty  $$
is then directly expressed in terms of the heat kernel.

More generally, if one considers a test function of the form $f_+(t)=e^{-|t|}$, then
for an operator $D=F \, |D|$ that satisfies a summability condition
\begin{equation}\label{sumLambda}
 \Tr(e^{-|D|/\Lambda}) < \infty 
\end{equation}
the spectral action $\Tr(f_+(D/\Lambda))$ with the test function as above
is well defined. Notice that, in such cases, there can be a ultraviolet
divergence problem if the condition \eqref{sumLambda} only holds for
energies below a certain scale, $\Lambda < \Lambda_0$. Notice that, 
if one imposes a cutoff $\Lambda < \Lambda_0$ on the energy scale,
then the test functions need not be smooth at the origin. 
In this setting, one can also consider a ``spectral asymmetry" version
of the spectral action, where instead of using an even test function one
uses an odd one. For instance one can consider $f_-(t)={\rm sign}(t)\, e^{-|t|}$
so that one obtains
\begin{equation}\label{sumoddLambda}
\Tr(f_-(D/\Lambda)) = \Tr( F e^{-|D|/\Lambda}),
\end{equation}
again with a cutoff $\Lambda \leq \Lambda_0$ for which the summabilty
condition \eqref{sumLambda} holds.

\subsection{Type III spectral triples} 

A variant on the notion of spectral triple was introduced recently by
Connes and Moscovici, \cite{CoMo3}, which uses a twisting by an
automorphism in the bounded commutator condition. This works
very well to extend the notion of spectral triple to certain
classes of noncommutative spaces that correspond to type III
von Neumann algebras, especially in the context of the geometry
of foliations.

We recall here the notion of a $\sigma$-spectral triple from \cite{CoMo3}.

\begin{defn}\label{sigmaSp3}
Let $\cA$ be a unital involutive algebra, represented as bounded operators
on a Hilbert space $\cH$ and let $D$ be a self-adjoint operator with compact
resolvent, densely defined on $\cH$. The data $(\cA,\cH,D)$ define a $\sigma$-spectal
triple if there is an automorphism $\sigma \in \Aut(\cA)$ such that
\begin{equation}\label{sigmaDa}
D a - \sigma(a) D \in \cB(\cH)
\end{equation}
is a bounded operator, for all $a \in \cA$. The $\sigma$-spectral triple $(\cA,\cH,D)$
is even if, moreover, there is a $\Z/2\Z$-grading $\gamma$ on $\cH$ such that
$[\gamma,a]=0$ for all $a\in \cA$ and $D\gamma+\gamma D=0$. The $\sigma$-spectral
triple is Lipschitz-regular if
$$ |D|\, a - \sigma(a) |D| \in \cB(\cH) $$
for all $a\in \cA$. 
\end{defn}

\section{Quantum Statistical Mechanical Systems}\label{QSMsec}

In this section we review the basic setting of Quantum Statistical Mechanical Systems
in Noncommutative Geometry. For more details, we refer the reader to \cite{BraRob},
\cite{CoMa-book}.

\begin{defn}\label{QSMdef}
A quantum statistical mechanical system consists of the following data:
\begin{itemize}
\item A separable unital $C^*$-algebra $\cA$;
\item A one-parameter family of automorphisms, namely a continuous
group homomorphism 
\begin{equation}\label{timeev}
\sigma: \R \to \Aut(\cA);
\end{equation}
\item A representation $\pi: \cA \to \cB(\cH)$ of $\cA$ as bounded operators
on a separable Hilbert space $\cH$
\item A densely defined self-adjoint operator $H$ on $\cH$ with $\Sp(H)\subset [0,\infty)$, 
such that
\begin{equation}\label{covrepH}
\pi(\sigma_t(a))=e^{itH} \pi(a) e^{-itH}, \ \ \ \forall a\in \cA, \ \ \forall t\in \R.
\end{equation}
\end{itemize}
 \end{defn}

\subsection{Time evolution and Hamiltonian}

The one parameter family of automorphisms \eqref{timeev} is the {\em time evolution}
of the quantum statistical mechanical system.
The datum of the operator $H$ implementing the time evolution as in \eqref{covrepH}
in general depends on the choice of the representation $\pi: \cA \to \cB(\cH)$, not only on
the data $(\cA, \sigma)$. The operator $H=H_{(\cA,\sigma,\pi)}$ is called the {\em Hamiltonian}
of the quantum statistical mechanical system. It is in general an unbounded, densely defined, 
operator. A representation $\pi: \cA \to \cB(\cH)$ is called {\em finite energy} if there is an
operator $H$ satisfying \eqref{covrepH}, with spectrum bounded below. This is equivalent,
up to an irrelevant translation, to the condition $\Sp(H)\subset [0,\infty)$ in the definition above.

\subsection{Partition function and critical temperature}

Given data as in Definition \ref{QSMdef} for a quantum statistical mechanical system,
the {\em partition function} of the Hamiltonian $H$ is defined as
\begin{equation}\label{ZetaH}
Z(\beta)= \Tr( e^{-\beta H}),
\end{equation}
where $\beta$ is a non-negative real variable, representing the inverse temperature
(up to multiplication by the Boltzmann constant). 

An analog of the summability conditions defined for spectral triples is the property
that the operator $e^{-\beta H}$ is of trace class, $\Tr( e^{-\beta H})<\infty$, for all
sufficiently large $\beta>>0$. The {\em critical inverse temperature} of the
quantum statistical mechanical system is then the value $\beta_c\geq 0$ such that
$\Tr( e^{-\beta H})<\infty$, for all $\beta > \beta_c$.

\subsection{KMS states}

A {\em state} on a unital $C^*$-algebra $\cA$ is a continuous linear functional
$\varphi: \cA \to \C$ that is normalized $\varphi(1)=1$ and satisfies a positivity
condition $\varphi(a^* a) \geq 0$, for all $a\in \cA$. States should be thought of
as the analogs of {\em measures} in the noncommutative world. Extremal states,
namely those states that cannot be decomposed as convex combinations of other
states, are an analog of {\em points} (Dirac measures) for noncommutative spaces.

In the case of a quantum statistical mechanical system, an especially interesting
class of states is given by the KMS equilibrium states. At a given inverse temperature
$\beta>0$, a state $\varphi$ on the algebra of observables $\cA$ is a KMS$_\beta$
state if there is a dense subalgebra $\cA^{an}\subset \cA$ of {\em analytic elements},
invariant under the time evolution $\sigma$ and such that, for all $a,b \in \cA^{an}$
\begin{equation}\label{KMS}
\varphi(ab) = \varphi(b \sigma_{i\beta}(a)).
\end{equation}
We refer the reader to \cite{BraRob} for more details on this and other
equivalent formulations of the KMS condition, and to the discussion in
Chapter 3 of \cite{CoMa-book} on how to interpret KMS states at $\beta=\infty$ 
(zero temperature states).

\subsection{Gibbs states}

In particular, among KMS states, one has Gibbs states of the form
\begin{equation}\label{stGibbs}
\varphi_\beta(a)= \frac{\Tr(\pi(a) e^{-\beta H})}{\Tr(e^{-\beta H})},
\end{equation}
with $\pi: \cA \to \cB(\cH)$ a representation of the algebra
of observable with Hamiltonian $H$ implementing the time
evolution. Gibbs states \eqref{stGibbs} exist under the summability
condition on the partition function
\begin{equation}\label{Zsumcond}
Z(\beta)=\Tr(e^{-\beta H}) < \infty.
\end{equation}
This typically happens in a certain range of inverse temperatures $\beta > \beta_c$.

Notice the similarity between this condition and the form \eqref{sumLambda}
of the spectral action, under similar summability conditions.
 
\section{The Riemann gas and type III spectral triples}\label{BCsec}

In this section we analyze a first example, where we compare quantum statistical
mechanical systems and spectral triples, in the setting of the Bost--Connes system
(\cite{BC}, see also Chapter 3 of \cite{CoMa-book}) and the Riemann gas 
(\cite{Jul}, \cite{Spec1}, \cite{Spec2}).

\subsection{The Bost--Connes system}
The Bost--Connes system is a quantum statistical mechanical system, introduced 
in \cite{BC}, with the following data:
\begin{itemize}
\item The algebra of observables is the unital $C^*$-algebra $C^*(\Q/\Z)\rtimes \N$,
with generators $\mu_n$, $n\in \N$ and $e(r)$, $r\in \Q/\Z$, and relations
$\mu_n^* \mu_n=1$; $\mu_n \mu_m =\mu_m \mu_n$; $\mu_n^* \mu_m =\mu_m \mu_n^*$ when $(n,m)=1$; $e(0)=1$; $e(r+s)=e(r)e(s)$ and 
\begin{equation}\label{BCrel}
 \mu_n^* e(r) = \sigma_n(e(r)) \mu_n^*, \ \ \  \mu_n e(r) = \rho_n(e(r)) \mu_n, \ \ \ 
 e(r) \tilde \mu_n =\tilde\mu_n \sigma_n(e(r)),
\end{equation} 
where $\sigma_n(e(r))=e(nr)$, with $\rho_n$ the partial inverse of $\sigma_n$, 
$$ \rho_n(e(r))=\frac{1}{n} \sum_{ns =r } e(s), $$
satisfying $\sigma_n \rho_n =id$ and $\rho_n \sigma_n =\pi_n$, the projection given by
the idempotent $\pi_n = \mu_n \mu_n^*=n^{-1} \sum_{ns =0} e(s)$.
\item The time evolution is given by $\sigma_t(e(r))=e(r)$, and $\sigma_t(\mu_n)=n^{it} \mu_n$,
for all $t\in \R$.
\item The Hilbert space $\cH =\ell^2(\N)$, with the standard orthonormal basis 
$\{ \epsilon_n \}_{n\in \N}$.
\item A Hilbert space representation, determined by a choice of an embedding 
$\alpha$ of the roots of unity in $\C$, given by 
$\mu_n\, \epsilon_m =\epsilon_{nm}$ (independently
of $\alpha$) and $\pi_\alpha(e(r)) \epsilon_n = \zeta_r^n \epsilon_n$, where $\zeta_r =\alpha(e(r))$
is a root of unity.
\item The Hamiltonian $H_\alpha$ is independent of the choice of $\alpha$
and given by $H \epsilon_n =\log(n)\, \epsilon_n$ and the partition function 
$\Tr(e^{-\beta H})=\zeta(\beta)$ is the Riemann zeta function. 
\end{itemize}
A complete classification of the KMS equilibrium states of the system was
given in \cite{BC}. We only recall here the fact that, in the low temperature range, namely
for $\beta >1$, the extremal KMS states are of the Gibbs form
\begin{equation}\label{GibbsBC}
\varphi_\beta(\pi_\alpha(e(r))) =\frac{\Tr(\pi_\alpha(e(r)) e^{-\beta H})}{\Tr(e^{-\beta H})}=
\frac{1}{\zeta(\beta)} \sum_{n\geq 1} \frac{\zeta_r^n}{n^\beta} = \frac{{\rm Li}_\beta(\zeta_r)}{\zeta(\beta)},
\end{equation}
where ${\rm Li}_s(z)$ is the polylogarithm function.

\begin{lem}\label{muLilem}
The Gibbs states \eqref{GibbsBC} can be equivalently written as 
\begin{equation}\label{muLi}
\varphi_\beta(\pi_\alpha(e(r))) =\sum_{n\geq 1} \frac{b_n}{n^\beta}, \ \ \text{ with } \ \ 
b_n = \sum_{d|n} \mu(\frac{n}{d}) \zeta_r^d,
\end{equation}
where $\zeta_r=\alpha(e(r))$, and $\mu$ is the M\"obius function.
\end{lem}

\proof  Using the relation
$$ \sum \frac{a_n}{n^\beta} = \zeta(\beta) \, \sum \frac{b_n}{n^\beta}, \ \ \  
\text{ with } \ \ \   b_n = \sum_{d|n}  \mu(\frac{n}{d}) \, a_d, $$
we rewrite the expression
$$ \varphi_\beta(\pi_\alpha(e(r))) = \frac{1}{\zeta(\beta)} \sum \frac{\zeta_r^n}{n^\beta} $$
as in \eqref{muLi}. 
\endproof

This quantum statistical mechanical system was extended from the case of $\Q$ to the
case of an arbitrary number field in \cite{CMR}, \cite{CorMa2}, \cite{HaPa}, 
\cite{LLN}, \cite{Yalk}.

\subsection{A type III spectral triple for the Bost--Connes algebra}

Recall that the Liouville function is defined as
\begin{equation}\label{Liouv}
\lambda(n) = (-1)^{\Omega(n)},
\end{equation}
where for an integer $n\in \N$ the number $\Omega(n)$ is the number of
prime factors of $n$, counted with the multiplicity. It satisfies 
$\lambda(nm)=\lambda(n)\lambda(m)$, since $\Omega$ is additive.

Consider the densely defined self-adjoint operator $D= F |D|$ on $\cH_B=\ell^2(\N)$, 
where $|D|=H$ is the Hamiltonian of the Bost--Connes system, 
$H \epsilon_n =\log(n)\, \epsilon_n$, and $F = (-1)^\Omega=\lambda$ is the bounded
linear operator on $\cH_B$ acting on the basis element $\epsilon_n$ as
multiplication by the Liouville function, $F \epsilon_n = \lambda(n) \epsilon_n$.
Consider also the operator $\tilde D = F |\tilde D|$ with the same sign $F$ and
with $|\tilde D|=\exp H$.

\begin{prop}\label{tilDetazeta}
The operator $\tilde D$ is finitely summable with
zeta function $\zeta_{\tilde D}(\beta)=Z(\beta)=\Tr(e^{-\beta H})=\zeta(\beta)$
equal to the partition function of the Bost--Connes system, and with eta
function
$$ \eta_{\tilde D}(\beta)= \Tr(F e^{-\beta H})= \frac{\zeta(2\beta)}{\zeta(\beta)}, $$
where $\zeta(\beta)$ is the Riemann
zeta function. The operator $D$ is $\theta$-summable with $\Tr(F e^{-\beta |D|})=\eta_{\tilde D}(\beta)$
and $\Tr(e^{-\beta |D|})=\zeta_{\tilde D}(\beta)$.
\end{prop}

\proof The result follows immediately, with the value of the eta function obtained through 
the well known formula
$$ \sum_{n\geq 1} \lambda(n) \, n^{-\beta} = \frac{\zeta(2\beta)}{\zeta(\beta)}, $$
where $\lambda$ is the Liouville function.
\endproof

We also introduce the following involutive automorphism of the 
Bost--Connes $C^*$-algebra.

\begin{lem}\label{sigmaOmega}
Let $\cA_\Q$ be the Bost--Connes $C^*$-algebra and let $\cA_\Q^{alg}\subset \cA_\Q$
be the dense involutive subalgebra generated algebraically by the elements $x\in \C[\Q/\Z]$
and the isometries $\mu_n$. The transformation $\sigma(a) = F a F$, for $a\in \cA_\Q$
is an automorphism of $\cA_\Q$ preserving $\cA_\Q^{alg}$. It satisfies $\sigma^2=1$
and it acts as the identity on the abelian subalgebra $C^*(\Q/\Z)$, while it acts
by $\sigma(\mu_n)=\lambda(n)\, \mu_n$ and $\sigma(\mu_n^*)=\lambda(n) \, \mu_n^*$
on the other generators.
\end{lem}

\proof It suffices to check that, if $a\in C^*(\Q/\Z)$, then $\sigma(a) =a$, while
for $a=\mu_n$ or $a=\mu_n^*$, $\sigma(a)=\lambda(n)\, a$. This is clear
since, in any representation $\pi_\alpha$, the operators $a\in C^*(\Q/\Z)$
act diagonally on the basis $\{ \epsilon_n \}$, hence they commute with $F$,
while $F \mu_n F \epsilon_\ell = \lambda(n\ell) \lambda(\ell) \epsilon_{n\ell}
=\lambda(n) \mu_n \epsilon_\ell$, and similarly in the case of $\mu_n^*$,
where $F \mu_n^* F \epsilon_\ell = \lambda(\ell/n) \lambda(\ell) \pi_n \epsilon_{\ell/n}=
\lambda(n)^{-1} \pi_n \epsilon_{\ell/n}=\lambda(n) \mu_n^* \epsilon_\ell$.
\endproof

Observe that the operator $H=|D|$ has the property that
the commutators $[|D|,a]$ are bounded operators, for all $a\in \cA_\Q^{alg}$.
In fact, $|D|$ commutes with elements $a\in \C[\Q/\Z]$, while, for the
generators $\mu_n$ and their adjoints one has
$$ [|D|,\mu_n] \epsilon_\ell = \log(n)\, \epsilon_{n\ell}, $$
and similarly for $\mu_n^*$. However, this is no longer the case
for $D=F|D|$, since, for instance, one has
$$ [D,\mu_n] \epsilon_\ell = ( \lambda(n\ell) \log(n\ell) - \lambda(\ell)\log(\ell)) \epsilon_{n\ell}, $$
which no longer gives a bounded operator when $\lambda(n)=-1$. However, we can
correct this problem by twisting the commutator relation, according to the prescription
described in \cite{CoMo} for type III $\sigma$-spectral triples. We obtain the following.

\begin{thm}\label{type3S3BC}
The data $(\cA_\Q^{alg}, \cH_B, D)$ determine a $\theta$-summable type III
$\sigma$-spectral triple, with respect to the automorphism $\sigma$ of Lemma \ref{sigmaOmega}.
This $\sigma$-spectral triple does not satisfy Lipschitz-regularity.
\end{thm}

\proof The operator $D$ is self-adjoint with compact resolvent by construction.
To see that it is $\theta$-summable, notice that the convergence of the series
$$ \Tr( e^{-\beta D^2})=\sum_{n\geq 1} e^{-\beta (\log n)^2} $$
is controlled by the integral
$$ \int_1^\infty e^{-\beta (\log x)^2}\, dx = \int_0^\infty e^{-\beta u^2+u}\, du \leq
\int_\R e^{-t^2 +\beta^{-1/2} t}\, dt = \sqrt{\pi} e^{1/(4\beta)}. $$ 

We then only need to check that the twisted commutator condition \eqref{sigmaDa}
of Definition \ref{sigmaSp3} is satisfied. We have
$D a - \sigma(a) D =0$ for $a\in \C[\Q/\Z]$ and 
$$ (D \mu_n - \sigma(\mu_n) D) \epsilon_\ell =
( \lambda(n\ell) \log(n\ell) - \lambda(n) \lambda(\ell)\log(\ell)) \epsilon_{n\ell} =
\lambda(n\ell) \log(n) \mu_n \epsilon_\ell, $$
which is a bounded operator. One shows similarly that, for a generating monomial
$a = x \mu_n \mu^*_m$, with $x\in \C[\Q/\Z]$ and $n,m\in \N$, 
the twisted commutator is bounded. Lipschitz-regularity fails because, for example,
$$ ( |D| \mu_n - \sigma(\mu_n) |D|) \epsilon_\ell = ( \log(n\ell)- \lambda(n) \log(\ell)) \epsilon_{n\ell}, $$
which is not a bounded operator when $\lambda(n)=-1$.
\endproof

Notice that, in general, when one sets $|D|=H$,
the convergence of the partition function of the quantum
statistical mechanical system may not always imply (nor be implied by)
the $\theta$-summable condition for the operator $|D|$. In fact, in general the
convergence of the partition function $\Tr(e^{-\beta H})$ will only happen 
for sufficiently low temperature $\beta > \beta_c$, where the critical inverse
temperature $\beta_c$ can be strictly positive, while the $\theta$-summable 
condition requires that $\Tr(e^{-t D^2})<\infty$ for all $t>0$.

\smallskip

We also have the following simple observation.

\begin{prop}\label{KMSsigma}
The functional $\varphi_\beta(a) = \Tr (F a e^{-\beta |D|})$ satisfies 
$$ \varphi_\beta(ab) = \varphi_\beta(b \sigma_{i\beta}(\sigma(a))), $$
for all $a,b\in \cA_\Q^{an}$, the analytic subalgebra of the Bost--Connes system,
with $\sigma_t$ the time evolution of the Bost--Connes system and $\sigma$
the automorphism of Lemma \ref{sigmaOmega}.
\end{prop}

\proof For $a\in \cA_\Q^{an}$ we have $e^{-\beta |D} a e^{\beta |D|}=\sigma_{i\beta}(a)$,
the analytic continuation to the strip of width $\beta$ of the time evolution $\sigma_t$.
Moreover, we have $\sigma_t \circ \sigma = \sigma \circ \sigma_t$,
since the sign $F$ commutes with $|D|$. We then have
$$ \varphi_\beta(ab) =\Tr(F\, ab\, e^{-\beta |D|}) = \Tr(F\, b (e^{-\beta |D|} F a F e^{\beta |D|})\, 
e^{-\beta |D|}) =  \varphi_\beta(b \sigma_{i\beta}(\sigma(a))). $$
\endproof

\subsection{The Riemann gas with supersymmetry}

Recall that the M\"obius function is defined as
\begin{equation}\label{Mobius}
\mu(n) =\left\{ \begin{array}{rl} +1 & n\ \text{squarefree with an even number of prime factors} \\
-1 & n\ \text{squarefree with an odd number of prime factors} \\
0 &  n\ \text{not squarefree.} \end{array}\right. 
\end{equation}
The vanishing for integers that contain squares, $p^2 | n$, can be seen, from a
physical perspective, as a Pauli exclusion principle (see \cite{Spec1}, \cite{Spec2}),
where one thinks of the primes as fermionic particles, where no two particles of
the same kind can occupy the same state. 

More precisely, the {\em bosonic} Fock space generated by the primes, as described
in \cite{BC}, is the tensor product $\cH_B = \bS \cH_P$, where $\cH_P=\ell^2(\cP)$,
with $\cP$ the set of primes, and $\cS\cH_P=\oplus_{n\geq 0} S^n \cH_P$ the Hilbert space
completion of the sum of the symmetric powers. 
This can be identified with $\cH_B=\ell^2(\N)$. The {\em fermionic}
Fock space, on the other hand, is defined as $\cH_F= \Lambda \cH_P =\oplus_n \Lambda^n \cH_P$,
the Hilbert space completion of the sum of the alternating powers. 

The Fermionic fock space has a canonical orthonormal basis given by the
elements $\{ p_1 \wedge \cdots \wedge p_n \, |\, p_i \in \cP \}$. Among these
basis elements, those containing an even number of primes ($n$ even)
are bosons, in the same way in which, in physics, a helium nucleus
is a boson, while the elements consisting of an odd number of primes ($n$ odd)
are fermions. 

As a Hilbert space, one can identify $\cH_F$ with the quotient of $\cH_B$ by the kernel
of the continuous linear functional $\mu: \cH_B \to \C$ defined on the canonical orthonormal
basis $\{ \epsilon_n \}_{n\in \N}$ of $\cH_B=\ell^2(\N)$ by the values of
the M\"obius function $\epsilon_n \mapsto \mu(n)$. The induced functional $\bar\mu: \cH_F \to \C$
on the quotient is a sign function $\bar\mu=(-1)^F$, which has values $+1$ on the bosons
and $-1$ on the fermions, hence it is the sign function that
implements Supersymmetry on $\cH_F$, as explained in \cite{Spec1}. Notice that
here one is implicitly using the fact that there is a natural choice of ordering of
the primes, namely the order in which they occur inside the natural numbers, see
\cite{Spec1} for explicit and more detailed comments on this point.

\subsection{A type III spectral triple for the Riemann gas}

As an example of how to relate the structures of spectral triples and quantum
statistical mechanical systems, we show here how to use the data of the 
Bost--Connes system and construct out of them a spectral triple related to
the supersymmetric Riemann gas. 

By the above description of the Hilbert spaces $\cH_B$ and $\cH_F$, we see that
we can identify $\cH_F$ with the subspace of $\cH_B$ spanned by those elements
$\{ \epsilon_n \}$ of the orthonormal basis of $\cH_B$ where $n$ is a squarefree integer.
We denote by $\Pi_F$ the orthogonal projection of $\cH_B$ onto this subspace,  
$\Pi_F: \cH_B \to \cH_F$.

In the following, we denote by $\cA_\Q$ the algebra of observables of the 
Bost--Connes system described above. We then 
consider the compression of $\cA_{\Q}$ by the projection $\Pi_F$. By this
we mean the algebra $\cA_{\Q,F} \subset \cB(\cH_B)$ generated by the operators
$\Pi_F \pi_\alpha(a) \Pi_F$, for all $a\in \cA_\Q$. This descends by construction
to an algebra $\cA_{\Q,F}\subset \cB(\cH_F)$ acting on the fermionic Hilbert space
$\cH_F$.

\begin{thm}\label{BCfermions}
The algebra $\cA_{\Q,F}$ is isomorphic to the $C^*$-algebra with generators 
$e(r)$, $r\in\Q/\Z$, and $\tilde\mu_n$, for $n\in \N$ squarefree, 
where the $e(r)$ satisfy the relations 
$e(s+r)=e(s)e(r)$ and $e(0)=1$, and generate $C^*(\Q/\Z)$ and the $\tilde\mu_n$
satisfy $\tilde\mu_n^* \tilde\mu_n=P_n$ and $\tilde\mu_n \tilde \mu_n^*=Q_n$, with
$P_n$ and $Q_n$ projections with $P_p+Q_p=1$, for $p$ a prime, and with 
$\tilde \mu_n \tilde\mu_m= \tilde \mu_m \tilde\mu_n =0$ if $(n,m)\neq 1$ 
and $\tilde\mu_{nm}$ if $(n,m)= 1$.
They also satisfy $\tilde \mu_m^* \tilde\mu_n =\tilde \mu_n \tilde \mu^*_m$ if $(n,m)=1$,
and 
\begin{equation}\label{emurel}
\tilde \mu_n^* e(r) = \sigma_n(e(r)) \tilde \mu_n^*, \ \ \  \tilde \mu_n e(r) = \rho_n(e(r)) \tilde \mu_n,
\ \ \  e(r) \tilde \mu_n =\tilde\mu_n \sigma_n(e(r)), 
\end{equation}
with $\sigma_n$ and $\rho_n$ as in \eqref{BCrel}.
\end{thm}

\proof Elements of the algebra $\cA_{\Q,F}$, seen as a subalgebra of $\cB(\cH_B)$, are
by construction of the form $\Pi_F \pi_\alpha(a) \Pi_F$, with $a\in \cA_\Q$. 
The algebra $\cA_\Q$ is the closure of the linear span of monomials of the 
form $\mu_n\, x\, \mu_m^*$, with $x\in C^*(\Q/\Z)$ and $n,m\in \N$, where 
$\mu_1=\mu_1^*=1$. First observe that, in the representations $\pi_\alpha \cA_\Q \to \cB(\cH_B)$
described above, the operators $\pi_\alpha(e(r))$ are diagonal, hence they satisfy
$\Pi_F \pi_\alpha(e(r)) = \pi_\alpha(e(r)) \Pi_F$. The relations \eqref{BCrel} for the algebra $\cA_\Q$
then imply that $\Pi_F \mu_n\, \pi_\alpha(x)\, \mu_m^* \Pi_F= \Pi_F \pi_\alpha(\rho_n(x)) 
\mu_n \mu_m^* \Pi_F=
\pi_\alpha(\rho_n(x)) \Pi_F \mu_n \mu_m^* \Pi_F$. Thus, $\cA_{\Q,F}$ is generated linearly
by elements of the form $\pi_\alpha(x)  \Pi_F \mu_n \mu_m^* \Pi_F$, with 
$x\in C^*(\Q/\Z)$ and $n,m\in \N$. The operator $\Pi_F \mu_n \mu_m^* \Pi_F$ acts
on a basis element $\epsilon_\ell$ of $\cH_B$ as zero, unless $\ell$ is squarefree,
$m|\ell$ (and in particular $m$ is itself square free), and $n \ell/m$ is squarefree.
If all these conditions are met, then $\Pi_F \mu_n \mu_m^* \Pi_F \epsilon_\ell = \epsilon_{n\ell/m}$.
Consider the operator $\Pi_F \mu_n \Pi_F \mu_m^* \Pi_F$. This acts on $\epsilon_\ell$ as zero,
unless $\ell$ is squarefree, $m|\ell$ and $\ell/m$ is squarefree,  $(n,\ell/m)=1$ and 
$n$ is squarefree. If these conditions are met then $\Pi_F \mu_n \Pi_F \mu_m^* \Pi_F \epsilon_\ell=
\epsilon_{n\ell/m}$. The condition that $\ell/m$ is squarefree is automatically satisfied since we
are already assuming that $\ell$ is. The condition $(n,\ell/m)=1$ with  
$n$ squarefree then implies that $n \ell/m$ is also squarefree, since $\ell/m$ is
squarefree. Conversely, if we know that $n \ell/m$ is squarefree, knowing that $\ell$
is squarefree, it follows that also $n$ and $\ell/m$ are square free and it also follows
that $(n,\ell/m)=1$. Thus, $\Pi_F \mu_n \mu_m^* \Pi_F=\Pi_F \mu_n \Pi_F \mu_m^* \Pi_F$,
hence $\cA_{\Q,F}$ is generated linearly by elements of the form
$\pi_\alpha(x)  \Pi_F \tilde\mu_n \tilde\mu_m^* \Pi_F$, where $\tilde\mu_n= \Pi_F \tilde\mu_n \Pi_F$
and $\tilde\mu_m^*=\Pi_F \tilde\mu_m^* \Pi_F$. The relations \eqref{BCrel} imply
that the elements $x\in C^*(\Q/\Z)$ and the generators $\tilde\mu_n$ satisfy the relations
\eqref{emurel}, since we have
$$ \tilde \mu_n^* \pi_\alpha(e(r)) = \Pi_F \tilde\mu_n^* \Pi_F \pi_\alpha(e(r)) = 
\Pi_F \tilde\mu_n^*\pi_\alpha(e(r))  \Pi_F =
\Pi_F \pi_\alpha(\sigma_n(e(r))) \mu_n^*  \Pi_F $$ 
$$= \pi_\alpha(\sigma_n(e(r)))  \Pi_F  \mu_n^*  \Pi_F =
\pi_\alpha(\sigma_n(e(r))) \tilde \mu_n^*, $$
and similarly for the other two relations in \eqref{emurel}. Moreover, we have
$\tilde \mu_n \tilde\mu_m \epsilon_\ell=0$ unless $n m \ell$ is squarefree
and equal to $\epsilon_{nm\ell}$ in that case.  The condition $n m \ell$ squarefree is
verified when so are $\ell$, $n$, and $m$ individually, with $(n,\ell)=1$, $(n,m)=1$
and $(m,\ell)=1$. Thus, when $(n,m)\neq 1$ we have $\tilde \mu_n \tilde\mu_m=0$
and when $(n,m)=1$, we have $\tilde \mu_n \tilde\mu_m=\tilde \mu_{nm}
=\tilde \mu_m \tilde\mu_n$, since the
latter acts on $\epsilon_\ell$ as zero unless $nm\ell$ is square free and as
$\epsilon_{nm\ell}$ when that is the case. The case of 
$\tilde \mu_m^* \tilde\mu_n =\tilde \mu_n \tilde \mu^*_m$ for $(n,m)=1$ is similar.
Observe then that $\tilde \mu_n^* \tilde\mu_n$ acts on $\epsilon_\ell$ as zero
unless $\ell$ is square free and $(n,\ell)= 1$, in which case it acts as the identity.
Thus $P_n$ is the projection onto the subspace of $\cH_F$ spanned by the
basis elements $\epsilon_\ell$ with such that none of the prime factors of $n$ divides $\ell$.
This is the same as the projection $\prod_{p|n} (1- \mu_p \mu_p^*) \Pi_F$, where the projections
$\Pi_F$ and $\prod_{p|n} (1- \mu_p \mu_p^*)$ commute, since the latter is an element of
$C^*(\Q/\Z)$.  The element $\tilde\mu_n \tilde\mu_n^*$ acts on $\epsilon_\ell$ as zero
unless $\ell$ is square free and $n$ divides $\ell$, in which case it is the identity. Thus,
for a single prime $p$, we have $Q_p =1-P_p$.
\endproof

Consider then the operator $\tilde D= F |\tilde D|$ with $F=\mu$ the operator on $\cH_B$
defined by the M\"obius function and $|\tilde D| = \exp H$, with $H$ the Hamiltonian of
the Bost-Connes system. Consider also the operator $D= F |D|$, with the same $F$ and
with $|D| = H$.

\begin{prop}\label{DHF}
The operators $F$ and $|\tilde D|$ commute and the operator $\tilde D$ defines
a densely defined unbounded self-adjoint operator on $\cH_F$ 
with trivial kernel and with compact resolvent, which is finitely summable.
The zeta function satisfies
\begin{equation}\label{zetaDtil}
 \zeta_{\tilde D}(\beta)=\Tr_{\cH_F}(|\tilde D|^{-\beta})=\frac{\zeta(\beta)}{\zeta(2\beta)}, 
\end{equation}
while the eta function satisfies
\begin{equation}\label{etaDtil}
\eta_{\tilde D}(\beta)=\Tr_{\cH_F}(F |\tilde D|^{-\beta})=\zeta(\beta)^{-1}, 
\end{equation}
where $\zeta(\beta)$
is the Riemann zeta function. The operator $D$ also defines an unbounded self-adjoint operator
on $\cH_F$, which is $\theta$-summable, with 
$\Tr_{\cH_F}(F e^{-\beta |D|})=\Tr_{\cH_F}(F |\tilde D|^{-\beta})$
and equal to the Witten index.
\end{prop}

\proof We have $F\epsilon_\ell = \mu(\ell) \epsilon_\ell$, with $\mu(\ell)$ the
value at the integer $\ell$ of the M\"obius function, and 
$|\tilde D| \epsilon_\ell =\ell \, \epsilon_\ell$. These satisfy $[F,|\tilde D|]=0$. 
Since $\tilde D$ preserves $\cH_F$, identified as a subspace of $\cH_B$ as
above, it induces an unbounded  self-adjoint linear operator
on this space, which satisfies the compact resolvent condition,
since $(1+\tilde D^2)^{-1/2} \epsilon_\ell = (1+\ell^2)^{-1/2} \epsilon_\ell$.
The zeta function is given by
$$ \zeta_{\tilde D}(\beta)= \Tr_{\cH_F}(|\tilde D|^{-\beta}) = 
\sum_{n\geq 1} |\mu(n)|\, n^{-\beta} = \frac{\zeta(\beta)}{\zeta(2\beta)}, $$
where the absolute value $|\mu(n)|$ of the M\"obius function is the
Dirichlet inverse of the Liouville function \eqref{Liouv}, 
while, by the M\"obius inversion formula, we have 
$$ \eta_{\tilde D}(\beta)=\Tr_{\cH_F}(F |\tilde D|^{-\beta})= \sum_{n\geq 1} \mu(n)\, n^{-\beta} = \frac{1}{\zeta(\beta)}. $$

The operator $D$ is theta summable and, 
as observed in \cite{Spec1}, the zeta function $\Tr_{\cH_F}(F |\tilde D|^{-\beta})$, which
is the same as $\Tr_{\cH_F}(F e^{-\beta |D|})$, computes the Witten index \cite{Witten}
\begin{equation}\label{Wittenind}
\Delta = \Tr((-1)^F \, e^{-\beta H}),
\end{equation}
where $(-1)^F$ is the sign on $\cH_F$ that implements the supersymmetry
(see also \cite{AkCom}, \cite{CecGir}).
\endproof

In particular, by Proposition \ref{DHF} the eta function \eqref{etaDtil}
is equal to the reciprocal of the Riemann zeta function, hence 
poles of the eta function $\eta_{\tilde D}(s)$ occur at zeros of the 
Riemann zeta function. This suggests interpreting the zeros as
phase transitions in this model.
An interpretation of zeros of zeta and $L$-functions as phase transitions
was also recently given in \cite{KiKoKu}.

\smallskip

We then obtain the following.

\begin{thm}\label{Sp3HF}
The data $(\cA_{\Q,F}, \cH_F, D)$, with $D$ as above, determine a
$\theta$-summable type III spectral triple, with respect to the automorphism
$\sigma\in \Aut(\cA_{\Q,F})$ given by $\sigma(a)=F a F$.
\end{thm}

\proof As a dense involutive subalgebra $\cA^{alg}_{\Q,F}$ of the $C^*$-algebra $\cA_{\Q,F}$
we can take the involutive algebra generated algebraically by the group
ring $\C[\Q/\Z]$ and the operators $\tilde\mu_n$, with the same relations as in
Theorem \ref{BCfermions}. The representation of $\cA_{\Q,F}$ on the Hilbert
space $\cH_F$ is induced by the choice of a representation $\pi_\alpha$ of
the algebra $\cA_\Q$ on $\cH_B$, together with the projection $\Pi_F$. The 
operator $D$ satisfies $[D, x]=0$ for all $x\in \C[\Q/\Z]$, as both operators 
are diagonal in the basis $\epsilon_\ell$, but we have 
$$ [D, \tilde\mu_n ] \epsilon_\ell = (\mu(n\ell) \log(n\ell) - \mu(\ell) \log(\ell)) \epsilon_\ell
=( \mu(n) \mu(\ell) \log(n) + (\mu(n)-1) \mu(\ell) \log(\ell)) \epsilon_{n\ell}, $$
where we used the fact that $(n,\ell)=1$ (otherwise $\mu_n\epsilon_\ell=0$)
and that for coprime integers the M\"obius function is multiplicative,
$\mu(n\ell)=\mu(n)\mu(\ell)$. Thus, we see that the commutators
$[D,a]$ are not always bounded, so the data $(\cA_{\Q,F}, \cH_F, D)$ do not
define an ordinary spectral triple. However, they do define a type III spectral
triple, in the sense of the $\sigma$-spectral triples defined in \cite{CoMo3}.
In fact, consider the automorphism $\sigma\in \Aut(\cA_{\Q,F})$ given by 
$\sigma(a)=F a F$. It is clearly an automorphism since $F^2=1$ and $F^*=F$, and 
it satisfies $\sigma^2=1$. For all elements $x\in C^*(\Q/\Z)$ this gives $\sigma(x)=x$,
while for the generators $\tilde \mu_n$ one has $\sigma(\tilde\mu_n)\epsilon_\ell
=F \tilde\mu_n \mu(\ell) \epsilon_\ell= \mu(n) \mu(\ell)^2 \epsilon_{n\ell}$ for 
$(n,\ell)=1$ and $n$ and $\ell$ squarefree, and zero otherwise, 
so that we have $\sigma(\tilde\mu_n) =
\mu(n)\, \tilde \mu_n$, multiplication by the value of the M\"obius function
at $n$. Similarly we have $\sigma(\tilde \mu_n^*)=\mu(n)^{-1} \tilde\mu_n^*
=\mu(n) \tilde\mu_n^*$. Thus, we obtain that, for instance,
$$ ( D \tilde\mu_n - \sigma(\tilde \mu_n) D ) \epsilon_\ell =
\mu(n)\mu(\ell) \log(n) \, \epsilon_{n\ell}, $$
(or zero if $(n,\ell)\neq 1$ or $n$ is not squarefree), which gives a bounded operator.
Similarly, for all elements $a\in \cA^{alg}_{\Q,F}$ we obtain that $D a - \sigma(a) D$
is bounded, using the fact that elements of this algebra are finite linear
combinations of elements of the form $x \tilde \mu_n \tilde \mu_m^*$. The boundedness
for these elements is checked as in the case of $\tilde \mu_n$ above.
The $\theta$-summable condition follows as in Theorem \ref{type3S3BC}.
\endproof

\begin{rem}\label{noLip2} {\rm 
The $\sigma$-spectral triple obtained in this way does not satisfy the
Lipschitz-regularity condition of \cite{CoMo3}, because the operators
$|D| a - \sigma(a) |D|$ are not always bounded. For instance
$$ ( |D| \tilde\mu_n - \sigma(\tilde\mu_n) |D| ) \epsilon_\ell =
(\log(n) + \log(\ell) (1-\mu(n)) ) \epsilon_{n\ell}, $$
for $(n,\ell)=1$ and $n$ and $\ell$ squarefree. }
\end{rem}

\subsection{Eta functions and polylogarithms}

We consider here the functionals of the form
\begin{equation}\label{etaaD}
 \varphi_\beta(a) = \Tr(F a e^{-\beta |D|}) = \eta_{a,\tilde D}(\beta),
\end{equation}  
for $a\in \cA_{\Q,F}$, with $F$ and $D=F |D|$ as in
Theorem \ref{Sp3HF} above and $\tilde D$ as in 
Proposition \ref{DHF}. We write
$$ \varphi_\beta(a) = \Tr(F a e^{-\beta |D|}) =  \sum_{n\geq 1} \mu(n) n^{-\beta} \, \langle \epsilon_n |\pi_\alpha(a)\, \epsilon_n \rangle. $$
By the previous considerations, it suffices to evaluate it on monomials of the form
$e(r) \tilde\mu_n \tilde\mu_m^*$ in $\cA_{\Q,F}$. We see that non-trivial values occur 
for elements of the form $e(r) \tilde\mu_n \tilde\mu_n^*$. This means that we can restrict
to evaluating \eqref{etaaD} for $a\in \C[\Q/\Z]$. We have
\begin{equation}\label{etaaDpolylog}
 \varphi_\beta(e(r))=\eta_{e(r),\tilde D}(\beta)=  \sum_{n\geq 1} \mu(n) \frac{\zeta_r^n}{n^\beta},
\end{equation}
which is a ``M\"obius inversion" of the polylogarithm function 
\begin{equation}\label{polylog}
{\rm Li}_\beta(\zeta) =  \sum_{n\geq 1} \frac{\zeta_r^n}{n^\beta}.
\end{equation}
at a root of unity $\zeta_r=\pi_\alpha(e(r))$. 

\bigskip

\section{Limit sets of Schottky groups and graphs: from spectral triples for QSM}

We discuss in this section another example where one can
relate the notions of spectral triple and of quantum statistical
mechanical system. We refer here to a construction of
spectral triples on the limit set of Schottky groups, used in
the reconstruction result for Riemann surfaces in \cite{CorMa1}
and to a similar version for graphs constructed in \cite{dJ},
which in turn has applications to Mumford curves with
$p$-adic uniformization.

\subsection{Spectral triples for limit sets and for graphs}

We recall here briefly the construction of spectral triples
for limit sets of Schottky groups and for graphs, as in \cite{CorMa1},
\cite{dJ}. We let $\Lambda$ denote either the limit set in $\P^1(\C)$ of
a Schottky group $\Gamma \subset {\rm PSL}_2(\C)$, or the
boundary $\partial X$ of the universal covering tree of
a finite, connected, unoriented graph with Betti number $g\geq 2$
and vertex valencies $\geq 3$. In the graph case we denote by $\Gamma$ the
free group of rank $g$ acting freely on the covering tree, with quotient the graph.

We consider the unital commutative $C^*$-algebra $\cA=C(\Lambda)$. This
is an $AF$ $C^*$-algebra, since $\Lambda$ is topologically a 
Cantor set. We also consider the involutive dense subalgebra
$\cA^\infty=C(\Lambda,\Z)\otimes_\Z \C$, the algebra of locally constant functions.
The Hilbert space $\cH$ is the completion of $\cA^\infty$ (as a vector space)
in the norm $\| f \|^2 =\int_\Lambda |f|^2\, d\mu$, where $\mu$ is the Patterson--Sullivan
measure on $\Lambda$. There is a natural filtration on $\cA^\infty$ by 
$\{ \cH^\infty_n \}_{n \geq 0}$, where $\cH^\infty_n$ is the finite dimensional linear span of the
characteristic functions $1_w = 1_{\Lambda(w)}$ of the clopen subset
$\Lambda(w)\subset \Lambda$ of infinite reduced words in the generators
of $\Gamma$ that start with the finite word $w$ of length $\leq n$.  
For the rest of this section we use the notation 
$P_n: \cH \to \cH^\infty_n$, the orthogonal projection. The Dirac
operator is given by
\begin{equation}\label{Dirac}
|D| = \lambda_0 P_0 + \sum_{n\geq 1} \lambda_n Q_n, 
\end{equation}  
where $Q_n = P_n - P_{n-1}$ and $\lambda_n = (\dim \cH^\infty_n)^3$.
The operator $|D|$ is finitely summable (in fact $1$-summable). 

Since the main aspect of the these spectral triples considered in \cite{CorMa1} and \cite{dJ} is
the associated family of zeta functions $\zeta_{a,D}(s)=\Tr( a\, |D|^{-s})$, the Dirac operator
only matters up to a sign, hence one considers only $|D|$. We will discuss the problem
of the sign in the following subsection.

\subsection{Spectral triples and the sign problem}

The discussion presented here on the sign of the Dirac operator applies to both
the spectral triples for limit sets of Schottky groups in \cite{CorMa1} and the
spectral triples of graphs of \cite{dJ}. In both cases the $C^*$-algebra  
$\cA$ of the spectral triple is equal to the set of continuous functions on 
what is topologically a Cantor set. 

We consider here the question of the existence of a $\Z /2\Z$-grading, 
compatible with the spectral triple, giving it an even structure. We show
that, for these spectral triples, one cannot find a compatible choice
of a non-trivial grading and a non-trivial sign $F$ for the Dirac operator. 

\begin{lem} \label{mult}
Any grading operator $\epsilon$ for the spectral triples on limits sets of 
Schottky groups or on graphs is a multiplication operator by a 
$\pm1$-valued function.
\end{lem}

\proof  We write here $\Lambda$ for either the limit set of the Schottky group or the boundary 
$\partial X$ of the covering tree of the graph.

A grading operator $\epsilon$ compatible with the spectral triple
commutes with the action of $\cA$ by multiplication on the Hilbert space $\cH = L^2(\Lambda, d \mu)$,
with $\mu$ the Patterson--Sullivan measure. It can be easily shown that any such operator $\epsilon$ must itself be a multiplication operator, and since $\epsilon ^2 = 1$, that $\epsilon$ is multiplication by a $\pm 1 $-valued function. That is, $\epsilon = 1_E - 1_{E^c}$, for some measurable set 
$E \subset \Lambda$. 

For a word/open set $\Lambda(w)$, let $a \in C(\Lambda)$ be the continuous function $1_w=1_{\Lambda(w)}$ and let $f \in L^2(\Lambda, d \mu)$ also be $1_w$. Then
\[
\epsilon a f(x) = \epsilon(1_w)(x)
\]
and
\[
a \epsilon f(x) = 1_w(x) \cdot \epsilon(1_w)(x).
\]
Since $\epsilon$ commutes with $a$ this implies that $\epsilon(1_w)(x) = 0$ whenever $x \notin w$. Then it follows easily that for any two words $w$, $v$, containing $x$, that
\[
\epsilon(1_w)(x) = \epsilon(1_v)(x).
\]
Therefore we may define $g(x)$ to be equal to $\epsilon(1_w)(x)$ for any word $w$ containing $x$. Now let $v$ be any finite word. Given $x \in \Lambda$, either $x$ is in $v$ or it is not. In either case, one can see that $\epsilon(1_v)(x) = g(x)1_v(x)$. Since the linear span of the characteristic functions $1_w$ are dense in $\cH$, this proves that $\epsilon$ is equal to multiplication by $g(x)$. Since $\epsilon^2 = {\rm id}$, we see that $g(x)$ must be $\pm 1$-valued.
\endproof

With respect to this grading, the $\Z/2\Z$-decomposition of $\cH = \cH ^+ \oplus \cH ^-$ is given by
\begin{align*}
\cH^+ &= L^2(E,d\mu) \\
\cH^- &= L^2(E^c,d\mu) \\
\end{align*}
We then show that neither $E$ nor $E^c$ contains an open set of $\Lambda$.

\begin{lem}
\label{noOpen}
Suppose $\epsilon = 1_E - 1_{E^c}$ is a grading defining the even structure on the spectral triple. 
For no open set $U \subset \Lambda$ does $E$ or $E^c$ contain almost every point of $U$.
\end{lem}

\proof We may restrict our attention to the open sets corresponding to words $w$, of finite length in the generators, as these open sets form a basis for the topology. We denote these open sets by $\Lambda(w)$. Suppose $E$ contained almost every point of the open set, $\Lambda(w)$. Now, for even spectral triples, given an idempotent $e \in A$, $eP : e \cH ^+ \rightarrow e \cH ^-$ defines a Fredholm operator, where $P$ denotes the restriction of the sign operator $F$ to $\cH^+$.

Let $e$ be the characteristic function $1_w=1_{\Lambda{w}}$, which is an idempotent in $\cA$. Then $e\cH ^- = \{ 0 \}$, and therefore, $eP$ has infinite-dimensional kernel which contradicts the fact that it is Fredholm. Similarly if $E^c$ contains almost every point of the open set, $w$, then $e \cH^+ = \{0 \}$, and so $eP$ would have infinite-dimensional cokernel which is again a contradiction.
\endproof

As we recalled above, the Dirac operator, in both the Schottky group and the graph spectral triple, 
is given by the formula
\[
|D| = \sum \lambda_n Q_n,
\]
where $Q_n$ is the orthogonal projection onto the finite-dimensional subspace 
$\cA_n \ominus \cA_{n-1}$, with $Q_0=P_0$.
The following is an easy yet important observation for our discussion
\begin{lem}\label{eigen}
A function $\psi \in \cH$ is an eigenvector for $|D|$ if and only if $\psi$ is in the range of $Q_n$ for some $n \in \N$. A vector, $\psi$, is an eigenvector of $D$, (whatever the sign $F$ may be) implies that $\psi$ is an eigenvector of $|D|$ and hence is in the range of $Q_n$ for some $n$. In particular, each eigenvector of $D$ is locally constant.
\end{lem}

Now we arrive at the main conclusion.

\begin{thm}\label{NoGradeLambda}
There is no choice of grading $\epsilon$ and non-trivial 
sign $F$ giving the graph spectral triple an even structure.
\end{thm}

\proof
Let $\psi$ be an eigenvector of $D$ with eigenvalue $\lambda$. Then since $D$ anticommutes with $\epsilon$, $\epsilon \psi$ is an eigenvector with eigenvalue $-\lambda$. However, for no $\psi \neq 0$ in the range of $Q_n$ is $\epsilon \psi$ locally constant, which contradicts lemma \ref{eigen}. The easiest way to see this is to look at a basis vector for the range of $Q_n$,
\[
\psi = c_k 1_{w_1} - c_1 1_{w_k},
\] 
where $w_1$, $w_k$ are distinct words of length $n$ and hence are disjoint.
Then 
\[
\epsilon \psi = c_k 1_{w_1 \cap E} - c_1 1_{w_k \cap E} + c_1 1_{w_k \cap E^c} - c_k 1_{w_1 \cap E^c}
\]
By lemma \ref{noOpen}, $\epsilon \psi$ is not locally constant on any neighborhood of any point contained inside $w_k$. This argument holds for any non-zero linear combination of such basis vectors and therefore we have shown that $\epsilon$ sends every non-zero vector in the range of $Q_n$ outside of $Q_n$.
\endproof

This has the following immediate consequence.

\begin{cor}
For no choice of sign $F$ is the Chern character of the graph spectral triple non-trivial.
\end{cor}

\proof
All $AF$ $C^*$-algebras have trivial $K_1$-group. Therefore, for there to be any hope of obtaining a non-trivial Chern character for the spectral triple, one must exhibit a $\Z /2\Z$-grading compatible with the spectral triple, giving it an even structure, so that its Chern character  pairs with $K_0(A)$, but
Theorem \ref{NoGradeLambda} shows that there is no non-trivial compatible choice of a 
$\Z /2\Z$-grading and a sign $F$ for the spectral triple.
\endproof

\subsection{From spectral geometry to quantum statistical system}

In comparing the structures of spectral triple and of quantum statistical mechanical
system recalled at the beginning of the paper, it seems natural to expect that the
operator $|D|$ of the spectral triple should be related to the Hamiltonian generating
the time evolution in the quantum statistical mechanical system. We have seen
this, in the reverse direction, in the case of the Riemann gas in the previous section,
where we used $|D|=H$ to obtain a $\theta$ summable $\sigma$-spectral triple
from a quantum statistical mechanical system. Here we start, instead, from the
operator $|D|$ of the spectral triple. If we follow the same intuition, we would
expect to be able to take $|D|$ itself as generator of the time evolution.  However,
this requires enlarging the algebra $\cA$ of the spectral triple, as explained below.

\subsection{Enlarging the algebra and time evolution}

Given a spectral triple $(\cA,\cH,D)$ as above, 
let $\cA_D$ be the $C^*$-subalgebra of $\cH$ generated by the elements of the
$C^*$-algebra $\cA$ and the spectral projections of the Dirac operator $D$. 

\begin{lem}\label{timeevAD}
For a spectral triple $(\cA,\cH,D)$, where $D$ has trivial kernel, 
the time one-parameter family of automorphisms $\sigma_t (a) =|D|^{it} a |D|^{-it}$
define a time evolution on the algebra $\cA_D$. This time evolution is by inner automorphisms.
\end{lem}

\proof By the spectral theorem for self-adjoint operators, 
for every $t\in \R$, the unitary operator $|D|^{it}$ is a bounded operator
contained in the $C^*$-algebra generated by the spectral projections of $D$. 
Thus, for any $a\in \cA_D$, the element $|D|^{it} a |D|^{-it}$ is again in $\cA_D$
and $t\mapsto \sigma_t(a)$ defines a time evolution by inner automorphisms.
\endproof

In the case of the spectral triples of \cite{CorMa1}, \cite{dJ}, the algebra $\cA_D$
is generated by the functions $f\in C(\Lambda)$ and the projections $Q_n$. Passing
from the algebra $\cA=C(\Lambda)$ of the spectral triple to the algebras $\cA_D$
ensures that the time evolution $\sigma_t (a) =|D|^{it} a |D|^{-it}$ maps the algebra
to itself. 

Let $I_m$ denote an inductively constructed orthonormal basis of 
$\cH^\infty_m\ominus \cH^\infty_{m-1}$, obtained by Gram--Schmidt orthonormalization.
Elements $\phi\in I_m$ are by construction elements in $\cA=C(\Lambda)$.

\begin{prop}\label{timevolQn}
In the case of the spectral triples of limit sets of Schottky groups and of graphs,
the time evolution $\sigma_t (a) =|D|^{it} a |D|^{-it}$ on $\cA_D$ is given
explicitly by
\begin{equation}\label{sigmatQn}
\sigma_t(a) \psi= \sum_{m\leq \max\{ n, \ell \}} (\frac{\lambda_m}{\lambda_n})^{it}
\, Q_m \phi \cdot \psi,
\end{equation}
for elements $a=\phi \in I_n-I_{n-1}$ and $\psi \in I_\ell-I_{\ell-1}$, with
$\lambda_n = (\dim \cH^\infty_n)^3$.
\end{prop}

\proof The time evolution is determined by its values on elements of
the form $a=\phi \in I_n-I_{n-1}$ for all $n\geq 0$. In turn $\sigma_t (\phi)$ is
determined by its values on the orthonormal basis of elements in $I_n -I_{n-1}$
for $n\geq 0$. For such elements we have
$$ \sigma_t (\phi) \psi =|D|^{it} \phi |D|^{-it} \psi = \lambda_n^{-it} |D|^{it} \phi \psi $$
$$ = \sum_{m \leq \max\{ n, \ell \}} \lambda_m^{it} \lambda_n^{-it} \sum_{\eta
\in I_m - I_{m-1}} \langle \eta, \phi \psi \rangle\, \eta 
= \sum_{m \leq \max\{ n, \ell \}}  (\frac{\lambda_m}{\lambda_n})^{it} 
 Q_m \phi \psi. $$
 \endproof

Thus, one obtains a quantum statistical mechanical system where
the algebra of observables is $\cA_D$, the Hilbert space is the same as for the
spectral triple, and the Hamiltonian is $H=\log |D|$. However, 
the fact that the time evolution constructed in this way is inner is not a 
desirable feature: we will describe a different relation between spectral triples
and quantum statistical mechanical systems for limit sets and graphs that
avoids this problem in \S \ref{QSMtoS3sec} below.

\subsection{Partition function, zeta functions and Gibbs states}

The quantum statistical mechanical system constructed above from
the data of a finitely summable spectral triple identifies zeta functions
and Gibbs states in the following way.

\begin{prop}\label{QSMZandzetas}
Let $(\cA,\cH,D)$ be the finitely summable spectral triple for limit sets of
Schottky groups or for graphs, as described above, and let $(\cA_D,\cH, H=|D|)$
be the quantum statistical mechanical system obtained as in Lemma \ref{timeevAD}.
The partition function of the QSM system is equal to the zeta function of the Dirac 
operator of the spectral triple and, more generally, the family of zeta 
functions $\zeta_{a,D}(\beta)$, normalized by $\zeta_D(\beta)$, 
correspond to the values $\varphi_\beta(a)$
of a Gibbs KMS state at inverse temperature $\beta$.
\end{prop}

\proof This is a direct consequence of the identification $H=\log |D|$, with
$$ \varphi_\beta(a) =\frac{ \Tr(a \, e^{-\beta H}) }{\Tr(e^{-\beta H})} =
\frac{\Tr(a |D|^{-\beta})}{\Tr(|D|^{-\beta})} = \frac{\zeta_{a,D}(\beta)}{\zeta_D(\beta)}. $$
\endproof

\section{Limit sets of Schottky groups and graphs: from QSM to spectral triples}\label{QSMtoS3sec} 

We present here a different way of relating spectral triples and quantum statistical
mechanical systems for limit sets of Schottky groups and for graphs, which works
more similarly to the case of the Riemann gas discussed earlier in the paper, namely
by proceeding from quantum statistical mechanical systems to (type III) spectral triples.

\subsection{Quantum statistical mechanical systems for Schottky groups and graphs}
We recall from \cite{CorMa3} that one can construct quantum statistical mechanical
systems associated to graphs. A very similar construction can be done in the case
of limit sets of Schottky groups as we describe briefly here (see also \cite{Coor}, \cite{Lott}).

The $C^*$-algebra of observables of the quantum statistical mechanical
system is in this case the noncommutative algebra $\cA=C(\Lambda)\rtimes \Gamma$,
where $\Lambda$ is either the limit set in $P^1(\C)$ of the Schottky group 
$\Gamma \subset {\rm PSL}_2(\C)$, or else the boundary $\Lambda=\partial T_X$ of the
universal covering tree $T_X$ of the graph $X$, acted upon by the free group $\Gamma$,
the fundamental group of the graph. Every element in the algebra can be obtained
as a limit of elements of the form $\sum_\gamma f_\gamma \, U_\gamma$, with
$f_\gamma \in C(\Lambda)$ and $U_\gamma$ the unitary operators corresponding
to the group elements $\gamma \in \Gamma$.

In the case of the Schottky group, let $T=T_\Gamma$ denote the Cayley graph of the
Schottky group, embedded in the hyperbolic space $\H^3$. Then the Schottky group
$\Gamma$ acts freely on its Cayley graph and the limit set $\Lambda$
of $\Gamma$ is also the boundary $\Lambda =\partial T_\Gamma$ of the tree $T_\Gamma$,
mirroring what happens in the graph case. 

The time evolution is constructed in \cite{CorMa3} as the one-parameter family of
automorphisms of $\cA$ given by
\begin{equation}\label{timeevCrossProd}
\sigma_t (\sum_\gamma f_\gamma(\xi) \, U_\gamma) =\sum_\gamma
e^{ it \, B(x_0,\gamma x_0,\xi) } \,\, f_\gamma(\xi)\, U_\gamma,
\end{equation}
where the points $x, y$ are vertices on the tree $T_X$ of the graph (respectively the
Cayley graph $T_\Gamma$), with $x_0$ a base point, and $B(x_0,\gamma x_0,\xi)$
is the Busemann function
\begin{equation}\label{Buse}
B(x_0,\gamma x_0,\xi) = \lim_{x \to \xi} d(x_0,x)-d(\gamma x_0, x) .
\end{equation} 


\subsection{KMS state and Patterson--Sullivan measure}

It is also known (see \cite{CorMa3} and \cite{Coor}, \cite{Lott}) that the quantum
statistical mechanical system with algebra $C(\Lambda)\rtimes \Gamma$ and
time evolution \eqref{timeevCrossProd} has a unique inverse temperature 
$\beta=\delta(\Lambda)$, which is the critical exponent for the Poincar\'e series of $\Gamma$
(the Hausdorff dimension of the limit set $\Lambda$), for which the set of KMS$_\beta$
states is non-empty. This set consists of a unique KMS state, given by integration with respect
to the normalized Patterson--Sullivan measure $\mu_{x_0}$
\begin{equation}\label{KMSunique}
\varphi_{\beta,x_0}(\sum_\gamma f_\gamma(\xi) \, U_\gamma) =\int_\Lambda f_1(\xi) \, d\mu_{x_0}(\xi).
\end{equation}

\subsection{Hamiltonian and time evolution}

We reformulate the time evolution \eqref{timeevCrossProd}
in terms of an explicit covariant realization given by a Hilbert space
representation of the algebra of observables and a Hamiltonian.

\begin{thm}\label{thmHBuse}
The $C^*$-algebra $\cA=C(\Lambda)\rtimes \Gamma$ acts by bounded operators
on the Hilbert space $\cH = L^2(\Lambda, d\mu)\otimes \ell^2(\Gamma)$, where
$\mu=\mu_{x_0}$ is the Patterson--Sullivan measure, with the action given by
\begin{equation}\label{actfUgamma}
 f U_\gamma \,\, h\otimes \gamma' = f \upsilon_\gamma(h) \otimes  \gamma \gamma' ,
\end{equation} 
where $\upsilon$ denotes the action of $\Gamma$ on $L^2(\Lambda,\mu)$ induced by
the action of $\Gamma$ on $\Lambda$. In this representation, the time evolution
is implemented by the Hamiltonian
\begin{equation}\label{HBuse}
H \,\, h(\xi) \otimes \gamma = B(x_0, \gamma x_0,\xi) \,\, h(\xi) \otimes \gamma .
\end{equation}
\end{thm}

\proof First we check that \eqref{actfUgamma} indeed determines a representation
of $\cA$ on the Hilbert space $\cH$. We have
$$ f_1 U_{\gamma_1}\,\, f_2 U_{\gamma_2} \,\, h\otimes \gamma' =
f_1 U_{\gamma_1}\,\, f_2 \upsilon_{\gamma_2} (h) \otimes  \gamma_2 \gamma'   = f_1 \upsilon_{\gamma_1} (f_2 \upsilon_{\gamma_2} (h)) \otimes \gamma_1
\gamma_2 \gamma' , $$
where the action $\upsilon_\gamma$ is given by
$$ \upsilon_\gamma (h)(\xi) = \rho_{\gamma,\mu}^{-1/2} \, h(\gamma^{-1} \xi) = \rho_{\gamma,\mu}^{-1/2} \, \alpha_\gamma(h), $$
where $\rho_\gamma(\mu)$ is the Radon--Nikodym derivative
$$ \rho_{\gamma,\mu} = \frac{d \gamma^*\mu}{d\mu}, $$
and we use the notation $\alpha_\gamma(h)=h(\gamma^{-1} \xi)$. Thus we obtain 
$$  f_1 U_{\gamma_1}\,\, f_2 U_{\gamma_2} \,\, h\otimes \gamma' = 
f_1 \upsilon_{\gamma_1} (f_2 \upsilon_{\gamma_2} (h)) \otimes \gamma_1
\gamma_2 \gamma' = f_1 \alpha_{\gamma_1}(f_2) \rho_{\gamma_1\gamma_2,\mu}^{-1/2}
\alpha_{\gamma_1\gamma_2}(h)  \otimes \gamma_1
\gamma_2 \gamma'  $$
$$  = f_1 \alpha_{\gamma_1}(f_2) \,\, \upsilon_{\gamma_1\gamma_2}(h)  \otimes \gamma_1
\gamma_2 \gamma' = f_1 \alpha_{\gamma_1}(f_2) U_{\gamma_1 \gamma_2} \,\, h\otimes \gamma' ,$$
where we have used the fact that, for all $\gamma, \tilde\gamma \in \Gamma$,
$$ \frac{d \tilde\gamma^*\mu}{d\gamma^* \mu} \, \frac{d \gamma^*\mu}{d\mu} =
\frac{d \tilde\gamma^*\mu}{d\mu}. $$

We consider then the operators $e^{it \, H} \,\, f U_\gamma \,\, e^{-it\, H}$, with $H$ as in
\eqref{HBuse}. These act on elements $h\otimes \gamma'$ of the Hilbert space $\cH$ as 
$$ e^{it \, H} \,\, f U_\gamma \,\, e^{-it\, H} \,\, h\otimes \gamma' =
e^{it \, H} f \upsilon_\gamma( e^{-it\, B(x_0, \gamma' x_0, \xi)} h(\xi) ) \otimes \gamma \gamma' $$
$$ = e^{it\, H} f(\xi) e^{-it\, B(x_0, \gamma' x_0, \gamma^{-1} \xi)} \upsilon_\gamma(h)(\xi)
\otimes \gamma \gamma' = e^{it (B(x_0, \gamma\gamma' x_0, \xi) - 
B(x_0, \gamma' x_0, \gamma^{-1} \xi))} f(\xi) \upsilon_\gamma(h)(\xi) \otimes \gamma \gamma' . $$
We then use the fact that the Busemann function satisfies 
$$ B(x,y,\xi)=-B(y,x,\xi), \ \  B(x,y,\xi)+B(y,z,\xi)=B(x,z,\xi), \ \  
B(\gamma x, \gamma y, \gamma \xi)=B(x,y,\xi) $$
to transform the above into
$$ e^{it \, H} \,\, f U_\gamma \,\, e^{-it\, H} \,\, h\otimes \gamma' = 
e^{it (B(x_0, \gamma\gamma' x_0, \xi) + B(\gamma \gamma' x_0, \gamma x_0,  \xi))} 
f U_\gamma\,\, h\otimes \gamma'  $$ $$ = e^{it B(x_0,\gamma x_0,\xi)}\, f U_\gamma \,\, 
h\otimes \gamma' = \sigma_t( f U_\gamma) \,\, h\otimes \gamma' . $$
\endproof

\subsection{A type III spectral triple for limit sets and graphs} 

We first introduce a suitable sign operator $F$ for the Dirac operator of the
spectral triple. We proceed by analogy to what we did in the case of the
Riemann gas and we consider an operator of the form
\begin{equation}\label{FLambda}
F \,\, h \otimes \gamma = (-1)^{\ell(\gamma)} \, h \otimes \gamma, 
\end{equation}
where $\ell: \Gamma \to \Z$ is a group homomorphism.  

\begin{lem}\label{FsignLambda}
The operator $F$ of \eqref{FLambda} is a sign operator for $D= F |D|$ with
$|D|=H$ of \eqref{HBuse}.
\end{lem}

\proof This follows by checking that $[|D|,F]=0$.
\endproof

We now show that the quantum statistical mechanical system described above determines
a type III $\sigma$-spectral triple, in the sense of \cite{CoMo3}.

\begin{thm}\label{3Sp3Lambda}
Consider the automorphism $\sigma(a)=F a F$ of the algebra $\cA=C(\Lambda)\rtimes \Gamma$,
with $F$ as in \eqref{FLambda},
and the representation of $\cA$ on $\cH=L^2(\Lambda,d\mu)\otimes \ell^2(\Gamma)$ as in
Theorem \ref{thmHBuse} above. We also consider the dense subalgebra $\cA^\infty=\cA^{alg}
=C(\Lambda)\rtimes^{alg} \Gamma$, algebraically generated by functions in $C(\Lambda)$ 
and the unitaries $U_\gamma$, $\gamma \in \Gamma$. We set $D=F |D|$ with $F$ as in
\eqref{FLambda} and $|D|=H$, as in \eqref{HBuse}. Then the data $(\cA^\infty, \cH, D)$ define
a $\sigma$-spectral triple. 
\end{thm}

\proof The map $a \mapsto \sigma(a)$ is an automorphism since $\ell: \Gamma \to \Z$ is
a group automorphism. In fact, $\sigma(a)=F a F$ acts as 
$$ \sigma(f U_\gamma)\,\, h \otimes \gamma' =F \, f U_\gamma\, F\,\, h \otimes \gamma' =
(-1)^{\ell(\gamma')} (-1)^{\ell(\gamma\gamma')}\, fU_\gamma\,\, h \otimes \gamma', $$
where $\ell(\gamma\gamma')= \ell(\gamma)+\ell(\gamma')$, so that we have 
$\sigma(ab)=\sigma(a)\sigma(b)$ and $\sigma(f)=f$ for $f\in C(\Lambda)$ and
$\sigma(U_\gamma)=(-1)^{\ell(\gamma)} U_\gamma$.
Elements in the dense involutive subalgebra $\cA^\infty=\cA^{alg}$ are finite
sums $\sum_\gamma f_\gamma U_\gamma$. Thus, it suffices to check the
commutator properties for monomials of the form $f U_\gamma$. We have, for $|D|=H$,
$$ ( |D|\, f U_\gamma - f U_\gamma\, |D| ) \, h \otimes \gamma' =
( B(x_0, \gamma\gamma' x_0, \xi) - B(x_0, \gamma' x_0, \gamma^{-1}\xi) ) \,
f U_\gamma\, h \otimes \gamma' = B(x_0, \gamma x_0, \xi) \, fU_\gamma\, h \otimes \gamma' , $$
while for $D=F |D|$ we have
$$ (D\, f U_\gamma - f U_\gamma\, D ) \, h \otimes \gamma' = (-1)^{\ell(\gamma')} 
( (-1)^{\ell(\gamma)} B(x_0, \gamma\gamma' x_0, \xi) - B(x_0, \gamma' x_0, \gamma^{-1}\xi) ) \,
f U_\gamma\, h \otimes \gamma' . $$
The twisted commutator satisfies
$$ (D f U_\gamma - \sigma(f U_\gamma) \, D ) \, h \otimes \gamma' = (-1)^{\ell(\gamma\gamma')} 
B(x_0, \gamma x_0, \xi) \, fU_\gamma\, h \otimes \gamma' . $$
\endproof

Notice that, for the purpose of reconstruction of graphs from the associated
quantum statistical mechanical systems, in \cite{CorMa3} one considers, instead
of $\cA^{alg}$, a non-involutive subalgebra of $\cA$, as in the case of the 
quantum statistical mechanical systems of number fields in \cite{CorMa2}.

\bigskip

\bigskip

\subsection*{Acknowledgment} The first author was supported for this work by a
Summer Undergraduate Research Fellowship at Caltech. The second author is
partially supported by NSF grants DMS-0901221, DMS-1007207, DMS-1201512, 
and PHY-1205440. The second author thanks MSRI for hospitality and support.


\begin{thebibliography}{99}
\bibitem{AkCom} R.~Akhoury, A.~Comtet, {\em Anomalous behavior of the Witten index
-- exactly soluble models}, Nucl. Phys. B, Vol.246 (1984) 253--278. 
\bibitem{BC} J.B.~Bost, A.~Connes, {\em Hecke algebras, Type III factors and phase transitions with
spontaneous symmetry breaking in number theory}, Selecta Math. Vol.1 (1995) N.3, 411--457. 
\bibitem{BraRob} O.~Bratteli and D.W.~Robinson, {\em Operator algebras and quantum statistical mechanics}, Vol.2, second ed., Texts
and Monographs in Physics, Springer-Verlag, 1997.
\bibitem{Cac} B.~\'Ca\'ci\'c, {\em A reconstruction theorem for almost-commutative spectral triples}, 
Lett. Math. Phys. 100 (2012), no. 2, 181--202.
\bibitem{CecGir} S.~Cecotti, L.~Girardello, {\em Functional measure, topology, and dynamical supersymmetry breaking}, Phys. Lett. B, Vol.110 (1982), 39--43.
\bibitem{CC1}  Ali Chamseddine, Alain Connes, {\em The spectral action principle}, Comm. Math. Phys. 186 (1997), no. 3, 731--750.
\bibitem{CCM} Ali Chamseddine, Alain Connes, Matilde Marcolli, {\em 
Gravity and the standard model with neutrino mixing}, 
Advances in Theoretical and Mathematical Physics, 11 (2007) 991--1090.
\bibitem{Co-hyp} A.~Connes, {\em Compact metric spaces, Fredholm modules, and 
hyperfiniteness}, Ergod. Th. Dynam. Sys. (1989) 9, 207--220.
\bibitem{Co94} A.~Connes, {\em Noncommutative Geometry}, Academic Press, 1994.
\bibitem{CoS3} Alain Connes, {\em Geometry from the spectral point of view}, 
Lett. Math. Phys. 34 (1995), no. 3, 203--238. 
\bibitem{CoCKM} A.~Connes, {\em 
A unitary invariant in Riemannian geometry}, Int. J. Geom. Methods Mod. Phys. 5 (2008), no. 8, 
1215--1242.
\bibitem{Co08} A.~Connes, {\em On the spectral characterization of manifolds}, 
J. Noncommutative Geom. Vol.7 (2013) N.1, 1--82.
\bibitem{CoMa-garden} A.~Connes, M.~Marcolli, {\em A walk in the noncommutative garden}, in ``An invitation to noncommutative geometry", pp. 1--128, World Scientific, 2008. 
\bibitem{CoMa-book} A.~Connes, M.~Marcolli, {\em Noncommutative Geometry,
Quantum Fields and Motives}, Colloquium Publications, Vol.55,
American Mathematical Society, 2008.
\bibitem{CMR} A.~Connes, M.~Marcolli, N.~Ramachandran, {\em KMS states and complex multiplication}, Selecta Math. (N.S.) 11 (2005), no. 3-4, 325--347.
\bibitem{CoMo} A.~Connes, H.~Moscovici, {\em The local index formula in noncommutative geometry}, Geom. Funct. Anal. 5 (1995), no. 2, 174--243.
\bibitem{CoMo3} A.~Connes, H.~Moscovici, {\em Type III and spectral triples}, in ``Traces in number theory, geometry and quantum fields", pp. 57--71, Aspects Math., E38, Vieweg, 2008.
\bibitem{Coor} M.~Coornaert, {\em 
Mesures de Patterson-Sullivan sur le bord d'un espace hyperbolique au sens de Gromov}, 
Pacific J. Math. 159 (1993), no. 2, 241--270. 
\bibitem{CordJ} G.~Cornelissen, J.W.~de Jong, {\em The spectral length of a map between 
Riemannian manifolds}, J. Noncommutative Geom. 6 (2012), no. 4, 721--748. 
\bibitem{CorMa1} G.~Cornelissen, M.~Marcolli, {\em Zeta Functions that hear the shape of a Riemann surface}, Journal of Geometry and Physics, Vol. 58 (2008) N.1 57--69.
\bibitem{CorMa2} G.~Cornelissen, M.~Marcolli, {\em Quantum statistical mechanics, $L$-series and anabelian Geometry}, preprint, arXiv: 1009.0736.
\bibitem{CorMa3} G.~Cornelissen, M.~Marcolli, {\em Graph reconstruction and quantum statistical mechanics}, preprint arXiv:1209.5783, to appear in Journal of Geometry and Physics.
\bibitem{Gil} P.B.~Gilkey, {\em Asymptotic formulae in spectral geometry}, Studies in Advanced Mathematics. Chapman Hall/CRC, 2004.
\bibitem{HaPa} E.~Ha, F.~Paugam, {\em Bost-Connes-Marcolli systems for Shimura varieties. I. Definitions and formal analytic properties}, IMRP Int. Math. Res. Pap. 2005, no. 5, 237--286.
\bibitem{HR} N.~Higson, J.~Roe, {\em Analytic $K$-homology},  Oxford University Press, 2000.
\bibitem{dJ} J.W.~de Jong, {\em Graphs, spectral triples and Dirac zeta functions}, 
p-Adic Numbers Ultrametric Anal. Appl. 1 (2009), no. 4, 286--296. 
\bibitem{Jul} B.~Julia, {\em Statistical theory of numbers}, in ``Number Theory and Physics", Springer, 1990.
\bibitem{KiKoKu} T.~Kimura, S.~Koyama, N.~Kurokawa, {\em 
Euler Products beyond the Boundary}, arXiv:1210.1216.
\bibitem{LLN} M.~Laca, N.~Larsen, S.~Neshveyev, {\em On Bost-Connes types 
systems for number fields}, J. Number Theory 129 (2009), no. 2, 325--338. 
\bibitem{LRV} S.~Lord, A.~Rennie, J.C.~Varilly, {\em Riemannian manifolds in noncommutative geometry},  J. Geom. Phys. 62 (2012), no. 7, 1611--1638.
\bibitem{Lott} J.~Lott, {\em Limit sets as examples in noncommutative geometry}, 
K-Theory 34 (2005), no. 4, 283--326.
\bibitem{Rieff} M.A.~Rieffel, {\em Compact quantum metric spaces}, in ``Operator Algebras, Quantization, and Noncommutative Geometry", Contemp. Math., Vol. 365, Amer. Math. Soc., 
2004, pp. 315--330.
\bibitem{Spec1} D.~Spector, {\em Supersymmetry and the M\"obius inversion function}, Commun. Math. Phys. Vol.127 (1990) 239--252.
\bibitem{Spec2} D.~Spector, {\em Duality, partial supsersymmetry, and arithmetic number theory}, J. Math. Phys. Vol.39 (1998) N.4, 1919--1927.
\bibitem{Witten} E.~Witten, {\em Constraints on supersymmetry breaking}, 
Nucl. Phys. B, Vol.202 (1982) 253--316. 
\bibitem{Yalk} B.~Yalkinoglu, {\em On arithmetic models and functoriality of Bost-Connes 
systems},  arXiv:1105.5022.
\bibitem{Zhang} D.~Zhang, {\em Projective Dirac operators, twisted K-theory and local index formula}, arXiv:1008.0707, to appear in J. Noncommutative Geom.

\end{thebibliography}
\end{document}